\documentstyle[11pt]{article}

\newtheorem{thm}{Theorem}[section]

\begin{document}

\title{\bf  {On Axially Symmetric  Space-Times Admitting   Homothetic Vector Fields  in Lyra's Geometry}}
\author{{ Ragab M. Gad$^{1,2}$ \thanks{%
E-mail: ragab2gad@hotmail.com}\,\, and A. E. Al Mazrooei$^{1}$\thanks{%
E-mail:abo.mhmd@hotmail.com } }\\
\newline
{\it $^1$
 Mathematics Department, Faculty of Science, University of Jeddah,}\\
 {\it  21589 Jeddah, KSA}
 \\
{\it $^2$ Mathematics Department, Faculty of Science, Minia University,}\\
 {\it   61915 El-Minia,  Egypt}
}

\date{\small{}}

\maketitle

\begin{abstract}
This paper investigates axially symmetric  space-times that admit  a homothetic vector field
   based on Lyra's geometry. The cases when the displacement vectors is a function
of $t$ and when it is constant are studied. In the context of this geometry, we find and classify the
solutions of the Einstein's field equations for the space-time under consideration, which display
a homothetic symmetry.
\end{abstract}

\setcounter{equation}{0}
\section{Introduction}

Space-time symmetries play an important role in the features of space-time that can be
described as exhibiting some form of symmetry. The most important of symmetries is in
simplifying Einstein field equations (EFE) and provide a classification of the space-times
according to the structure of the corresponding Lie algebra.\\

Symmetries have been studied in the theory of general relativity (GR) based on Riemannian
geometry and in the theory of teleparallel gravity  based on the Weitzenbock geometry.
In GR different kinds of symmetries like isometry,
homothetic, conformal, Ricci  collineations and matter collineations have  been
extensively studied \cite{H04}-\cite{G03}.
Also some of them have been studied in teleparallel gravity, over the
past few years \cite{SM09}-\cite{S11}.\\
In a series of papers, Gad and Alofi \cite{G1}-\cite{G3} studied the homothetic
symmetry based on Lyra's geometry. They classified the space-times according to admitting such
symmetry. For zero displacement
vector field, they obtained results that agree with those obtained previously in GR, based on Riemannian geometry.
They showed also that  in the case when the displacement vector field is constant, it is not possible
to compare the results obtained in the context of Lyra's geometry with that obtained in GR, using Riemannian geometry.
This means that in Lyra's geometry if the displacement vector field is considered as
a constant this does not give meaningful results.\\
In the framework of Lyra's geometry, many authors have made attempts
to find solutions of EFE \cite{P09}- \cite{S08}. These attempts were made
by imposing certain conditions upon the scale factors of the
space-time in addition to the conditions on the matter that
represents such space-time and some restrictions on its physical
properties. In the same context, we will obtain solutions to EFE
for an axially symmetric space-time by assuming only that such
space-time displays a homothetic symmetry, that is, exhibits a
self-similarity of the first kind (cf. \cite{CT71,{HW78},{BH78}}).

The paper is organized as follows: In the next Section, we will study the homothetic
symmetry of axially symmetric  space-times  based on  Lyra's geometry.  Section 3 deals with the solutions of EFE
 and their classifications of the space-times under considerations. Finally, in Sect. 4,
  concluding remarks are given.

\setcounter{equation}{0}
\section{\bf{ Version of model and homothetic vector field in Lyra's geometry}}
In Lyra's geometry the metric or the measure of length of displacement vector
$\zeta^\mu =x^o dx^\mu$ between two points $p(x^\mu)_{\mu=1}^n$ and $q(x^\mu+dx^\mu)_{\mu=1}^n$
is given by absolute invariant under both gauge function,  $x^o =x^o(x^\mu)_{\mu=1}^n$,
and coordinate system, $\{x^\mu\}_{\mu=1}^{n}$, as follows
 $$
 ds^2=g_{\mu\nu}x^odx^\mu x^o dx^\nu,
 $$
 where $g_{\mu\nu}$ is a metric tensor as in Riemannian geometry.\\
In Lyra's geometry, a generalized affine connection  characterized not only by the Riemannian connection,
$\big\{\,^\alpha_{\mu\nu}\big\}$, but also by a function $\phi_\mu$, which arises through gauge transformation,
  is given by
\begin{equation}\label{conn}
\Gamma^{\alpha}_{\,\,\mu\nu}=(x^o)^{-1}\big\{\,^\alpha_{\mu\nu}\big\} + \frac{1}{2}\big(\delta^\alpha_\mu\phi_\nu +
\delta^\alpha_\nu\phi_\mu -g_{\mu\nu}\phi^\alpha\big),
\end{equation}
where   $\phi$ is called a displacement vector field and satisfies $\phi^\alpha =g^{\alpha\beta}\phi_\beta$.
 We consider $\phi$ to be a time-like vector, where
\begin{equation}\label{phi}
\phi_\mu=(\beta(t),0,0,0).
\end{equation}

Throughout the paper $M$ will denoted a $4$-dimensional Lyra manifold with Lorentz metric
$g$ which is a generalization to the $4$-dimensional Riemannian manifold \cite{L51,S52}.
As in Riemannian geometry, a global vector field ${\bf{\zeta}}=\zeta^\mu(t,x,y,z)_{\mu=1}^4$
on $M$ is called {\it{homothetic vector field}} if the following condition
\begin{equation}\label{HE}
\pounds_\zeta g_{\mu\nu}=g_{\rho\nu}\nabla_\mu\zeta^\rho +g_{\mu\rho}\nabla_\nu\zeta^\rho=2\psi g_{\mu\nu},
\end{equation}
holds where $\psi$ is a constant (the {\it{homothetic constant}}) on $M$, $\pounds$ denotes a
 Lie derivative and $\nabla$ is the covariant derivative, such that
\begin{equation}
\begin{array}{ccl}
\nabla_\mu\zeta^\nu &=&\frac{1}{x^o}\partial_\mu\zeta^\nu + \Gamma^{\nu}_{\,\,\mu\alpha}\zeta^\alpha,\\
\nabla_\mu\zeta_\nu &=&\frac{1}{x^o}\partial_\mu\zeta_\nu - \Gamma^{\alpha}_{\,\,\mu\nu}\zeta_\alpha.
\end{array}
\end{equation}
Here $\Gamma^{\sigma}_{\mu\nu}$ is a Lyra connection given by equation (\ref{conn}).\\
In equation (\ref{HE}), if $\psi \neq 0$,  $\zeta$ is called {\it{ proper homothetic}} vector field and the this equation is called {\it{homothetic equation}}. It is worth mentioning here that  if $\psi =0$, then equation (\ref{HE}) is called {\it{Killing equation}} and $\zeta$ is called a {\it{Killing vector field}} on $M$.

Consider the axially symmetric metric  in the form \cite{BK93}
\begin{equation}\label{met}
ds^2 = dt^2 -A^2(t)(d\chi^2 +f^2(\chi) d\phi^2) -B^2(t)dz^2,
\end{equation}
with the convention $x^0=t$, $x^1=\chi$, $x^2=\phi$, $x^3=z$ and  $A$ and $B$ are
functions of $t$ only while $f$ is a function of the coordinate $\chi$ only.\\

The study of homothetic vector fields, ${\bf{\zeta}}=\zeta^\mu(t,x,y,z)_{\mu=1}^4$, on axially symmetric space-times (\ref{met}) is based on an examination of the 10 equations obtained from (\ref{HE}). For the model (\ref{met}), using (\ref{conn}) and apart from the factor $\frac{1}{x^o}$, (i.e., we choose the normal gauge $x^o=1$)  (\ref{HE}) are reduced to the
following system of equations:
\begin{equation}\label{1}
\zeta^1_{,\chi} + \big(\frac{A_{,t}}{A} +\frac{1}{2}\beta\big)\zeta^0=\psi,
\end{equation}
\begin{equation}\label{2}
\zeta^1_{,y}+f^2\zeta^2_{,\chi}=0,
\end{equation}
\begin{equation}\label{3}
A^2\zeta^1_{,z}+B^2\zeta^3_{,\chi}=0,
\end{equation}
\begin{equation}\label{4}
\zeta^0_{,\chi}-A^2\zeta^1_{,t}=0,
\end{equation}
\begin{equation}\label{5}
\zeta^2_{,y} +\frac{f_{,\chi}}{f}\zeta^1+\big(\frac{{A_{,t}}}{A} +\frac{1}{2}\beta\big)\zeta^0=\psi,
\end{equation}
\begin{equation}\label{6}
A^2f^2\zeta^2_{,z} +B^2 \zeta^3_{,y}=0,
\end{equation}
\begin{equation}\label{7}
\zeta^0_{,y}-A^2f^2\zeta^2_{,t}=0,
\end{equation}
\begin{equation}\label{8}
\zeta^3_{,z} +\big(\frac{{B_{,t}}}{B} +\frac{1}{2}\beta\big)\zeta^0=\psi,
\end{equation}
\begin{equation}\label{9}
\zeta^0_{,z}-B^2\zeta^3_{,t}=0,
\end{equation}
\begin{equation}\label{10}
\zeta^0_{,t} + \frac{1}{2}\beta\zeta^0=\psi.
\end{equation}

Solving  (\ref{10}) and using the result back into  (\ref{4}), (\ref{7}) and (\ref{9}), we get
\begin{equation} \label{1-}
\zeta^0=[\psi\int{e^{\frac{1}{2}\int{\beta dt}}dt} +c_0]e^{-\frac{1}{2}\int{\beta dt}},
\end{equation}
$$
\zeta^1=F_1(\chi,y,z),
$$
$$
\zeta^2=F_2(\chi,y,z),
$$
$$
\zeta^3=F_3(\chi,y,z),
$$
where $c_0$ is a constant of integration and $F_1(\chi,y,z)$, $F_2(\chi,y,z)$ and $F_3(\chi,y,z)$ are arbitrary functions  that are to be determined.\\
Differentiating equations (\ref{1}) and (\ref{8}) with respect to $t$ and using (\ref{10}) and (\ref{1-}), we get respectively
\begin{equation} \label{2-}
\frac{{A_{t}}}{A}+\frac{1}{2}\beta=\frac{a}{[\psi\int{e^{\frac{1}{2}\int{\beta dt}}dt} +c_0]e^{-\frac{1}{2}\int{\beta dt}}},
\end{equation}
\begin{equation} \label{2--}
\frac{{B_{t}}}{B}+\frac{1}{2}\beta=\frac{c}{[\psi\int{e^{\frac{1}{2}\int{\beta dt}}dt} +c_0]e^{-\frac{1}{2}\int{\beta dt}}},
\end{equation}

where $a$ and $c$  are constants of integration. Without loss of generality, we assume that $a=c$.  Substituting these results back into (\ref{1}) and (\ref{8}), using the obtained results in  (\ref{2}), (\ref{3}),  and (\ref{6}), we get
\begin{equation} \label{3-}
\zeta^1=(\psi -a)\chi +c_1,
\end{equation}
\begin{equation} \label{4-}
\zeta^3=(\psi -a)z +c_3,
\end{equation}
where $c_1$ and $c_3$ are constants of integration. Consequently, $\zeta^2$ depends on $y$ only.\\
Integrating  (\ref{2-}) and (\ref{2--}), using $a=c$, we get the following relation:
\begin{equation} \label{AB}
B(t)= nA(t),
\end{equation}
where $n$ is a constant of integration

From  (\ref{5}), we have the following two cases:
\begin{enumerate}
\item {\bf{Case i.}} If we assume that\\
\begin{equation}
\frac{f_{,\chi}}{f}\zeta^1=c_4\neq 0.
\end{equation}
Integrating this equation, we get
\begin{equation}\label{f1}
f(\chi)=c_5\chi^{\frac{c_4}{\psi-a}},
\end{equation}
where $c_5$ is a constant of integration.\\
Inserting the above results into (\ref{5}), we get
 \begin{equation} \label{6-}
\zeta^2=(\psi -b)y +c_2,
\end{equation}
where $c_2$ is a constant of integration and $b=a+c_4$.\\
Without loss of generality, we assume that $c_0 =c_1 =c_2 =c_3= 0$, therefore, from  (\ref{1-}), (\ref{3-}), (\ref{4-}) and (\ref{6-}) we obtain the following homothetic vector field
\begin{equation} \label{HVF}
{\bf{\zeta}}=([\psi\int{e^{\frac{1}{2}\int{\beta dt}}dt} ]e^{-\frac{1}{2}\int{\beta dt}})\, \partial_t + (\psi - a)\chi\,\partial_\chi +(\psi-b) y\partial_y +(\psi-a) z\partial_z.
\end{equation}
It is of interested to note that the function (\ref{f1}) does not satisfy that $\frac{f_{,\chi\chi}}{f}$ equals constant or zero. Consequently,  metric (\ref{met}) with $g_{22}=c_5\chi^{\frac{c_4}{\psi-a}}$ is not a solution to EFE, as we see in the next section. Therefore, this function will withdraw from consideration except the special case, when $c_4=\psi-a$. The latter case gives
$$
\zeta^2=c_2
$$
and
\begin{equation}\label{f2}
f(\chi)=c_5\chi.
\end{equation}
In this case, the homothetic vector becomes
\begin{equation} \label{HVF-}
{\bf{\zeta}}=([\psi\int{e^{\frac{1}{2}\int{\beta dt}}dt} ]e^{-\frac{1}{2}\int{\beta dt}})\, \partial_t + (\psi - a)\chi\,\partial_\chi +c_2 y\partial_y +(\psi-a) z\partial_z.
\end{equation}

\item {\bf{Case ii.}} If\\
\begin{equation}
\frac{f_{,\chi}}{f}\zeta^1= 0.
\end{equation}
- When $\zeta^1 \neq 0$, this implies
\begin{equation} \label{f3}
f(\chi)=c_6,
\end{equation}
where $c_6$ is a constant of integration.\\
In this case the homothetic vector is the same as given by  (\ref{HVF}).\\
When $\zeta^1=0$, this implies
\begin{equation} \label{f4}
f(\chi)=c_8e^{c_7\chi},
\end{equation}
where $c_7$ and $c_8$ are constants of integration.\\
In this case the homothetic vector is given by
\begin{equation} \label{HVF3}
{\bf{\zeta}}=([\psi\int{e^{\frac{1}{2}\int{\beta dt}}dt} ]e^{-\frac{1}{2}\int{\beta dt}})\, \partial_t +(\psi-b) y\partial_y +(\psi-c) z\partial_z.
\end{equation}

\end{enumerate}

All the preceding considerations lead to the following theorem:
\begin{thm}
In Lyra's geometry, if a displacement vector is function of $t$, that is, $\beta=\beta(t)$,
axially symmetric space-time described by metric (\ref{met}) admits the homothetic vector field  if
$$
B(t)= nA(t).
$$
Such homothetic vector field  is given by  (\ref{HVF}) if $f(\chi)$ is given by   (\ref{f3}) and  is given by  (\ref{HVF-}) or (\ref{HVF3}) if $f(\chi)$ is given by  (\ref{f2}) or (\ref{f4}), respectively.
\end{thm}

Now we will investigate the situation when the displacement vector is constant, that is, $\beta=$ constant.\\
In this case, from  (\ref{1-}), the time-like component of the homothetic vector $\zeta^0$, the only component depends on $\beta$, becomes
$$
\zeta^0 = (\frac{2\psi}{\beta}e^{\frac{1}{2}\beta t} +c_0)e^{-\frac{1}{2}\beta t}.
$$
While the space-like components are given as the case when $\beta=\beta(t)$ according to the values of $f(\chi)$.

Now we conclude that the homothetic vector field given by (\ref{HVF}) and with the above constrains on the scale factors  is  similar to the vector as obtained in the theory of GR (see for instance, \cite{Be12,Be13}).\\
 It is noteworthy that, in the case $\beta=$ const., we cannot compare the results with that obtained in the theory of GR, using Riemannian geometry,  because in this case the component $\zeta^0$  tends to infinity when $\beta=0$. This means that in Lyra's geometry if the displacement vector field is considered as a constant this does not give meaningful results.\\

\setcounter{equation}{0}
\section{Field equations and their solutions}
In this section, we shall determine the exact solutions of EFE by assuming that the space-time under consideration admits a homothetic vector field (self-similarity).

The field equations in normal gauge for Lyra's geometry as obtained
by Sen \cite{Sen57} (in gravitational units $c=8\pi G=1)$ read as
\begin{equation}\label{EFE-}
R_{\mu\nu}-\frac{1}{2}Rg_{\mu\nu}=-T_{\mu\nu}-\frac{3}{2}\phi_{\mu}\phi_{\nu}+
\frac{3}{4}g_{\mu\nu}\phi_{\alpha}\phi^{\alpha},
\end{equation}
the left hand side is the usual Einstein tensor as in Riemannian geometry, whereas
$\phi_{\mu}$ is a time-like displacement field vector defined by (\ref{phi})
and $T_{\mu\nu}$ is the energy momentum tensor corresponding to perfect
fluid.
It is  interesting to assume that the matter  field in space-time under consideration is represented by a perfect fluid,
 that is, the energy-momentum tensor is defined by
\begin{equation}\label{EMT}
T_{\mu\nu}=(\rho +p)u_{\mu}u_\nu-pg_{\mu\nu},
\end{equation}
Here $p$ is the pressure, $\rho$ the energy density and $u_{\mu}$ the
four velocity vector, it must verify $\pounds_{\zeta_\mu}u_{\mu}=0$. For the space-time (\ref{met}) the 4-velocity vector can be defined by $u^\mu=u_{\mu}=(1,0,0,0)$ and it is verified $g_{\mu\nu}u^\mu u^\nu=1$.
In view of the metric (\ref{met}), the volume element, the four-acceleration vector, the
rotation, the expansion scalar and the shear scalar  can be written, respectively, in a co-moving coordinates system as (see \cite{Glyra})

\begin{equation}\label{vol}
V = \sqrt{-g}= A^2fB,
\end{equation}

\begin{equation}\label{Kin-v}
\begin{array}{ccc}
\dot{u}_i & =& 0,\\
\omega_{ij} & = & 0,\\
\Theta & = & \frac{2{A_{,t}}}{A}+\frac{{B_{,t}}}{B},\\
\sigma^2 & = & \frac{1}{9}\big(11\big(\frac{{A_{t}}}{A}\big)^2+5\big(\frac{{B_{t}}}{B}\big)^2 + \frac{2{A_{t}}{B_{t}}}{AB}\big).
\end{array}
\end{equation}
The non-vanishing components of the shear tensor$\sigma_{ij}$ are
\begin{equation}\label{comp}
\begin{array}{ccc}
\sigma_{11} & =  & A(\frac{1}{3}\Theta A-{A_{,t}}) ,\\
\sigma_{22} & = & Af^2(\frac{1}{3}\Theta A-{A_{,t}}) ,\\
 \sigma_{33}& = & B(\frac{1}{3}\Theta B-{B_{t}}),\\
 \sigma_{44}& =  & -\frac{2}{3}\Theta.
\end{array}
\end{equation}

For the line element (\ref{met}), the field equations (\ref{EFE-})
with  (\ref{EMáT})  lead to the following system of equations (see for instance Gad \cite{Glyra})
\begin{equation}\label{E1}
\frac{{A_{,tt}}}{A} + \frac{{B_{,tt}}}{B} + \frac{{A_{,t}}{B_{t}}}{AB}
+ \frac{3}{4}\beta^2 = -p,
\end{equation}
\begin{equation}\label{E3}
\frac{2{A_{tt}}}{A} + (\frac{{A_{,t}}}{A})^2 -
\frac{f_{,\chi\chi}}{fA^2} +\frac{3}{4}\beta^2 =
-p,
\end{equation}
\begin{equation}\label{E4}
 (\frac{{A_{,t}}}{A})^2 +\frac{2{A_{t}}{B_{t}}}{AB}
 - \frac{f_{,\chi\chi}}{fA^2}-\frac{3}{4}\beta^2 =
 \rho,
\end{equation}
\begin{equation}\label{E5}
 {\rho_{,t}}+(\rho + p)(\frac{2{A_{t}}}{A}+\frac{{B_{t}}}{B})=0,
\end{equation}

In GR, Cahill and Taub \cite{CT71} and Bicknell and Henriksen [15] (see also \cite{OP90}) pointed out, if
the matter field is a perfect fluid, then the only barotropic equation of state that is
compatible with self-similarity (characterized by the existence of homothetic vector field) is of the form
\begin{equation}\label{bara}
p=k\rho,
\end{equation}
where $\rho$ is the total energy density, $p$ is the pressure and $k$ is a constant in the range
$0\leq k \leq 1$. This equation of state is nevertheless physically consistent in the whole range
of $k$. When $k = 0$, the preceding equation describes dust, $k =1/3$ gives the equation of state for
radiation and $k=1$ considers the effective "stiff fluid"
distribution. The latter was apparently first proposed by
Zeldovich \cite{Z62}. It would have applied in the early Universe,
because in this case, the
velocity of sound equals the velocity of light, so no material in
this Universe could be more stiff.\\

Exact solutions  for EFEs (\ref{E1})-(\ref{E5}) can be found under the assumption that the space-time (\ref{met}) admits homothetic vector field. The obtained  solutions can be classified according  to the value of the scale factor $f(\chi)$ and the constant $k$.\\
Using (\ref{AB}) and (\ref{bara}), the EFEs (\ref{E1})-(\ref{E5}) reduce to the following equations:
\begin{equation}\label{E-1}
\frac{2{A_{,tt}}}{A} + \big(\frac{{A_{,t}}}{A}\big)^2
+ \frac{3}{4}\beta^2 = -k\rho,
\end{equation}
\begin{equation}\label{E-2}
3\big(\frac{{A_{,t}}}{A}\big)^2
- \frac{3}{4}\beta^2 = \rho,
\end{equation}
\begin{equation}\label{E-3}
\rho=\frac{m}{A^{3(1+k)}},
\end{equation}
where $m$ is a constant of integration.\\
From  (\ref{E-1})-(\ref{E-3}), we get
\begin{equation}\label{E-4}
\frac{2{A_{tt}}}{A} + 4\big(\frac{{A_{t}}}{A}\big)^2
 = \frac{m(1-k)}{A^{3(1+k)}}
\end{equation}

In the following we discuss two cases, when $k=1$ and $k=0$:\\
\underline{\bf{Class I:}}\,\,\, k=1\\
In this case
\begin{equation}\label{A1}
A(t)=(\alpha t+\alpha_1)^{\frac{1}{3}},
\end{equation}
where $\alpha$ and $\alpha_{1}$ are constants of integration.\\
 According to the values of $f(\chi)$, given by (\ref{f2}), (\ref{f3}) and (\ref{f4}), we have, respectively, the following solutions
\begin{equation}\label{met1}
ds^2 = dt^2 -(\alpha t+\alpha_1)^{\frac{2}{3}}(d\chi^2 +c_5\chi d\phi^2 +ndz^2),
\end{equation}
\begin{equation}\label{met2}
ds^2 = dt^2 -(\alpha t+\alpha_1)^{\frac{2}{3}}(d\chi^2 +c_6 d\phi^2 +ndz^2),
\end{equation}
\begin{equation}\label{met3}
ds^2 = dt^2 -(\alpha t+\alpha_1)^{\frac{2}{3}}(d\chi^2 +c_8e^{c_7\chi} d\phi^2 +ndz^2),
\end{equation}
\underline{\bf{Class II:}}\,\,\, $k=0$\\
From  (\ref{E-4}), we have the following equation
\begin{equation}\label{E-5}
A^2{{A_{,tt}}} + 2A{{A_{,t}^2}}
 = \frac{m}{2}.
\end{equation}
Solving (ref{E-5}), we get
\begin{equation}\label{A2}
A(t)=\frac{1}{2}(6m t^2-\alpha_2 t+\alpha_3)^{\frac{1}{3}}.
\end{equation}
where $\alpha_2$ and $\alpha_{3}$ are constants of integration.\\
For the values of $f(\chi)$, as given in theorem 2.1, we have the following class of solutions
\begin{equation}\label{met4}
ds^2 = dt^2 -\frac{1}{4}(6m t^2-\alpha_2 t+\alpha_3)^{\frac{2}{3}}(d\chi^2 +c_5\chi d\phi^2 +ndz^2),
\end{equation}
\begin{equation}\label{met5}
ds^2 = dt^2 -\frac{1}{4}(6m t^2-\alpha_2 t+\alpha_3)^{\frac{2}{3}}(d\chi^2 +c_6 d\phi^2 +ndz^2),
\end{equation}
\begin{equation}\label{met6}
ds^2 = dt^2 -\frac{1}{4}(6m t^2-\alpha_2 t+\alpha_3)^{\frac{2}{3}}(d\chi^2 +c_8e^{c_7\chi} d\phi^2 +ndz^2),
\end{equation}

For the  class I of solutions, the expressions for density $\rho$,
pressure $p$  and displacement field $\beta$ are given by
 $$
 p = \rho= \frac{m}{(\alpha t+ \alpha_1)^2},
 $$
 which shows that $\rho$ and $p$ are not singular,
 $$
 \beta^2 = \frac{4}{9}\big(\frac{\alpha^2-3m}{(\alpha t+ \alpha_1)^2}\big).
 $$
It is observed, from  (\ref{A1}) and (\ref{AB}) that $A(t)$ and $B(t)$ can be singular only for $t \rightarrow \infty$.
Thus the line element (\ref{met}) is singularity-free even at $t=0$.\\

For the same class of solutions, using  (\ref{vol})-(\ref{comp}), we have the following physical properties:\\
The volume element is
$$
V =nf(\chi) (\alpha t+\alpha_1).
$$
Here $f(\chi)$ takes one of the values given by  (\ref{f2}), (\ref{f3}) and (\ref{f4}). For all values of $f(\chi)$, we see that the volume element increases as the time increases. This show that the solutions (\ref{met1})-(\ref{met3}) are expanding with time.\\

The expansion scalar, which determines the volume behavior of the fluid, is given by
$$
\Theta = \frac{\alpha}{\alpha t+\alpha_1}.
$$
The only non-vanishing component of the shear tensor, $\sigma_{ij}$, is
$$
\sigma_{44}=- \frac{2\alpha}{3(\alpha t+\alpha_1)}.
$$
Hence, the shear scalar $\sigma$ is given by
$$
\sigma^2 =2\big(\frac{\alpha}{9(\alpha t+\alpha_1)}\big)^2.
$$

Because $\lim_{t\rightarrow\infty} (\frac{\sigma}{\Theta})\neq 0$, then the solutions (\ref{met1})-(\ref{met3}) do not approach isotropy
for large value of $t$ and do not admit  acceleration and rotation, because $\dot{u}_i =0$ and $\omega_{ij}=0$.\\

As the preceding discussion,    class II of solutions (\ref{met4})-(\ref{met6}) has the following physical properties
$$
p=0,
$$
 $$
  \rho= \frac{8m}{6m t^2-\alpha_2 t+\alpha_3},
 $$
 $$
 \beta^2 = \frac{4}{9}\big(\frac{\alpha_2^2-24m\alpha_3}{(6m t^2-\alpha_2 t+\alpha_3)^2}\big),
 $$
 $$
 V=\frac{n}{4}f(\chi)(6m t^2-\alpha_2 t+\alpha_3),
$$
$$
\Theta = \frac{12mt-\alpha_2}{6m t^2-\alpha_2 t+\alpha_3},
$$
$$
\sigma_{44}=- \frac{2}{3}\big(\frac{12mt-\alpha_2}{6m t^2-\alpha_2
t+\alpha_3}\big),
$$
$$
\sigma^2 =\frac{2}{9}\big(\frac{12mt-\alpha_2}{6m
t^2-\alpha_2t+\alpha_3}\big)^2,
$$
and
$$
\lim_{t\rightarrow\infty} (\frac{\sigma}{\Theta})\neq 0.
$$

\section{Discussion and conclusion}
Homothetic symmetry is  one of the most important types of symmetries, because it plays  a dominant role in the
 dynamics of cosmological models \cite{RJ85,C03}.\\
 In this paper, we studied this symmetry of axially symmetric  space-times  within the framework of Lyra's geometry.  When the displacement vector, {\bf{$\phi$}}, is function of $t$, we obtained homothetic vector field and found that its expression depends on the scale factor $f(\chi)$. In the case when {\bf{$\phi$}} is constant the time component of the homothetic vector field tends to infinity when $\beta=0$. This means that we cannot compare the obtained results with those obtained in GR, using Riemannian geometry.\\
 The second aim of this paper was to use the homothetic symmetry to simplify  EFE. We assumed that the space-time under consideration admits this symmetry and found the exact solutions without more assumptions on the space-time as made in  most of the literatures. We classified the obtained solutions  according to the values of $f(\chi)$ and the constant $k$ into two classes. We found that the two classes of solutions are singularity-free at the initial epoch $t=0$ and have vanishing accelerations. For these solutions $\lim_{t\rightarrow \infty}(\frac{\sigma}{\Theta}) \neq 0$,  that is, they do not approach isotropy for large time $t$. All obtained solutions are  expanding with time because their  volume element increases as the time increases.

\section*{Acknowledgments}
This project was funded by the Deanship of Scientific Research (DSR), University of Jeddah, Jeddah, under grant No. (G-1436-965-348).
The authors, therefore, acknowledge with thanks DSR technical and financial support.


\begin{thebibliography}{99}

\bibitem{H04} G. S. Hall , Symmetries and curvature structure in general relativity. World
     Scientific. (2004).
\bibitem{SKMHH03} H. Stephani, D. Kramer, M. A. H. MacCallum, C. Hoenselears,  and  E. Herlt. Exact
     solutions of Einstein's field equations. Cambridge University Press. (2003).
 \bibitem{SM07} G. Shabbir   and  A. B.  Mehmood.  Mod. Phys. Lett.  A, {\bf{22}}, 807 (2007).
\bibitem{H96} G. S. Hall. Grav. Cosmol., {\bf{2}}, 270 (1996).
\bibitem {H98} G. S. Hall. Gen. Rel. Grav., {\bf{30}}, 1099 (1998).
\bibitem{CT71} M. E. Cahill  and A. H. Taub. Commun. Math. Phys., {\bf{21}}, 1 (1971).
\bibitem{E74} D. M. Eardley. Commun. Math. Phys., {\bf{37}}, 287 (1974).
\bibitem{E74b} D. M. Eardley. Phys. Rev. Lett., {\bf{33}},  442  (1974).
\bibitem{CH91} B. Carter   and R. N. Henriksen. J. Math. Phys., {\bf{32}}, 2580  (1991).
\bibitem{SBC01} A. M. Sintes, P. M. Benotit   and A. A. Coley. Gen. Rel. Grav., {\bf{33}}, 1863  (2001).
\bibitem{W00} J. Wainwright. Gen. Rel. Grav., {\bf{32}}, 1041  (2000).
\bibitem{W78} P. S. Wesson. J. Math. Phys., {\bf{19}},  2283  (1978).
\bibitem{CL87} M. E. Collins   and J. M.  Lang. Class. Quantum Grav., {\bf{4}},  61 (1987).
\bibitem{S98}  R. A. Sussman   J. Math. Phys., {\bf{32}}, 223  (1991).
\bibitem{BOR98}   W. Barreto, J.  Ovalle   and  B.  Rodriguez. Gen. Rel. Grav., {\bf{30}}, 15  (1998).
\bibitem{OP90}  A. Ori  and T. Piran. Phys. Rev. D, {\bf{42}}, 1068 (1990).
\bibitem{G02}  R. M. Gad.  Il Nuovo Cimento  {\bf{117B}}, 533  (2002).
\bibitem{G02}  R. M. Gad.  Il Nuovo Cimento  {\bf{124B}}, 61  (2009). doi:10.1393/ncb/i2009-10745-3.
\bibitem{G03} R. M. Gad  and M. M. Hassan. Il Nuovo Cimento {\bf{118B}}, 759  (2003). doi:10.1393/ncb/i2003-10055-x.


\bibitem{SM09} M. Sharif  and  B. Majeed. commun. Theor. Phys. {bf{52}}, 435 (2009).
\bibitem{SK10}  G. Shabbir    and S. Khan. Mod. Phys. Lett. A, {\bf{25}},  55  (2010).
\bibitem{SK10b} G. Shabbir    and S. Khan. Mod. Phys. Lett. A, {\bf{25}},  525  (2010).
\bibitem{SK10c} G. Shabbir    and S. Khan. Mod. Phys. Lett. A, {\bf{25}}, 1733  (2010).
\bibitem{SK10d} G. Shabbir    and S. Khan. Commun. Theor. Phys., {\bf{54}},  469  (2010).
\bibitem{SKA11} G. Shabbir    and S. Khan. and  A. Ali. Commun. Theor. Phys., {\bf{55}}, 268  (2011).
\bibitem{SA08} M. Sharif   and   M. J. Amir. Mod. Phys. Lett. A, {\bf{23}}, 963  (2008).
\bibitem{S11}   G. Shabbir, A. Ali   and  S. Khan. Chin. Phys. B, {\bf{20}}, 070401  (2011).


\bibitem{G1} R. M.  Gad  and A.S. Alofi. Mod. Phys. Lett. A, {\bf{22}}, 1450116 (2014).
\bibitem{G2} R. M. Gad . Int. J. Theor. Phys., {\bf{54}},  2932 (2015)
\bibitem{G3}  A. S. Alofi  and  R. M. Gad. Canad. J. Phys {\bf{93}}, 1397 (2015).



\bibitem{P09} A. Paradhan. J. Math. Phys. {\bf{50}}, 022501 (2009)
\bibitem{}  R. Casana, C. Melo,  and B. Pimentel.  Astrophys. Space Sci.{\bf{305}}, 125 (2006).
\bibitem{} F.  Rahaman, B. Bhui,  and G. Bag. Astrophys. Space Sci.{\bf{295}}, 507 (2005).
\bibitem{} R. Bali  and N. K. Chandani.   J. Math. Phys. {\bf{49}}, 032502
(2008)
\bibitem{} S. Kumar  and C. P. Singh.  Int. J. Mod. Phys., {\bf{A23}}, 813 (2008).
\bibitem{} V. U. M. Rao, T. Vinutha,  and M. V. Santhi. Astrophys. Space Sci.{\bf{314}}, 213 (2008).
\bibitem{S08} J. K. Singh. Astrophys. Space Sci. {\bf{314}}, 361 (2008).



\bibitem{HW78}  R. N. Henriksen,  and P. S. Wesson. Astrophys. Space Sci., {\bf{53}}, 429 (1978).
\bibitem{BH78} G. V. Bicknell,  and R. N. Henrikson. Astrophs. J., {\bf{219}},1043 (1978).


\bibitem{L51} G. Lyra.  Math. Z. {\bf{54}}, 52 (1951).
\bibitem{S52} E. Scheibe.  Math. Z. {\bf{57}}, 65 (1952).


\bibitem{BK93} S. Bhattacharaya,  and T. M. Karade. Astrophys. Space Sci.{\bf{202}}, 69 (1993).


\bibitem{Be12} J. A. Belinchon. Cent. Eur. J. Phys. {\bf{10}}, 789 (2012).
\bibitem{Be13}  J. A. Belinchon. Astrophys. Space Sci. {\bf{346}}, 237 (2013).


\bibitem{Sen57} D. K. Sen. Z. Phys. {\bf{149}}, 311 (1957).

\bibitem{Glyra} R. M. Gad. Can.J. phys., {\bf{89}}, 773 (2011).


\bibitem{Z62} Ya. B. Zeldovich. Sov. Phys. JETP {\bf{14}}, 1143 (1962).



\bibitem{RJ85} R. Rosquist,  and R. Jantzen. Class. Quantum Gravity {\bf{2}}, L129 (1985).
\bibitem{C03} A. A. Coly. Dynamical System and Cosmology. Kluwer Academic, Dordrech (2003).



\end{thebibliography}
\end{document}